\documentclass[11pt]{article}

\usepackage[letterpaper,margin=0.85in]{geometry}
\usepackage{amsmath}
\usepackage{amssymb}
\usepackage{graphicx}
\usepackage[round,authoryear]{natbib}
\usepackage{microtype}
\usepackage[colorlinks=true,linkcolor=blue,citecolor=blue,urlcolor=blue]{hyperref}
\graphicspath{{./}{figures/}}
\providecommand{\bibinfo}[2]{#2}

\title{Xenia: A Table-Driven Code for Astrophysical Multi-Fluid Hydrodynamics}
\author{Lei Xing\,\href{https://orcid.org/0000-0001-7655-0920}{\raisebox{-0.32ex}{\includegraphics[height=2.0ex]{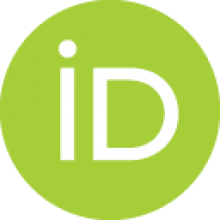}}}\thanks{Email: \href{mailto:yfyg4669@outlook.com}{yfyg4669@outlook.com}; ORCID: \href{https://orcid.org/0000-0001-7655-0920}{0000-0001-7655-0920}.}\thanks{Beijing ENN Fusion Energy Science and Technology Co., Ltd., Beijing 101111, China.}\thanks{Beijing Key Laboratory of High Magnetic Field Spherical Torus Fusion Energy; Hebei Key Laboratory of Compact Fusion.}}
\date{}

\begin{document}

\maketitle

\begin{abstract}
Models of escaping planetary atmospheres require neutral and ion species with distinct velocities and thermal pressures, coupled by photoionization, recombination, radiation, and collisional exchange. Implementing and extending such coupled models remains a practical barrier because species definitions and stiff interactions are often embedded in solver-specific kernel code. We present Xenia, a table-driven code for astrophysical multi-fluid hydrodynamics designed to lower that barrier. Its initial application surface is transonic, radiation-coupled planetary atmospheric escape. Fluid species, coupling channels, chemistry, heating and cooling, and radiation channels are declared in Python, compiled to symbolic tapes, and passed across a versioned interface to shared C/Kokkos kernels. The code assembles the coupled source terms from versioned tables, so changing a problem changes data rather than kernel code. For ionized configurations, electrons provide a pressure-only quasi-neutral closure: electron density follows from ion populations, no electron mass or momentum is advanced, and only electron pressure is transported. Stiff coupling is treated by an operator-split sequence of radiation scan, hyperbolic transport, and cell-local implicit source solve; the same symbolic tapes yield an analytic finite-difference-free Jacobian for a direct dense LU solve. Validation uses the archived v0.1 release to cover acoustic convergence, analytic dustybox relaxation, conservation and closure contracts, two-fluid cloud--shock coupling, curvilinear geometry bookkeeping, three escape chains, and 394 passing tests. Conservation and positivity are not claimed to match mature production codes; radiation is a radial Beer--Lambert prefix product rather than a general solver; escape chains provide reproducibility checks rather than new science predictions. Backend evidence is limited to functional deployment and equivalence fingerprints. The public code release is available at \href{https://github.com/YFG-tec/xenia}{https://github.com/YFG-tec/xenia} under GPLv2.
\end{abstract}

\noindent\textbf{Keywords:} methods: numerical --- hydrodynamics --- planetary atmospheres --- exoplanets --- software

\section{Introduction}

Problems such as the escape of planetary atmospheres and the
interaction between magnetospheres and the solar wind share a
common modeling requirement: neutral and ionized species --- for
example hydrogen, helium, and their ions --- each carry their own
velocity and thermal pressure, while electrons in ionized
configurations provide a quasi-neutral pressure closure.  These species are
coupled by photoionization, recombination, and collisional exchange. A
single bulk flow with a
fixed composition captures neither the fractionation between
species --- helium separating from hydrogen in an outflow, for
instance --- nor the heating and cooling set by this photochemistry.
On the escape side, the requirement runs through the literature:
atmospheric escape from hot Jupiters \citep{murrayclay2009},
simulations of hot-Jupiter upper atmospheres \citep{trammell2014},
the dehydration of water worlds through atmospheric losses
\citep{dong2017}, and helium atmospheres around Earth-mass planets
in the habitable zone \citep{lammer2025}. On the space-weather
side, distinct ion populations must likewise carry their own
momentum and thermal pressure. Of these problem classes, the
escape of planetary atmospheres is the primary science driver of
the code presented here, and transonic, radiation-coupled
escaping-atmosphere models supply the initial suite of problems
it is built to run.

Our own entry point was the escaping atmosphere of HD 209458b,
modeled with multi-fluid hydrodynamics including helium mass
fractionation \citep{xing2023}, followed by a two-fluid model of
the escaping atmosphere of GJ 436 b \citep{xing2024}. Both were
built on a private layer carved out of a heavily modified PLUTO
\citep{mignone2007pluto} --- an arrangement that produced the
science but offered no reusable framework.

The literature uses ``multi-fluid'' for at least three inequivalent
model classes; conflating them obscures what a code actually
solves. We define each in turn and state which one this paper
means.

The first meaning is multi-component or reactive-network modeling:
several chemical species are tracked as abundance fields of a
single bulk flow, sharing one velocity and one pressure throughout,
as in FLASH \citep{fryxell2000flash},
RAMSES \citep{teyssier2002ramses}, and the chemistry module
distributed with PLUTO \citep{mignone2012pluto}.

The second meaning is a gas plus pressureless dust: one thermally
active fluid is coupled by drag to one or more dust species that
have velocities of their own but carry no pressure, as in
FARGO3D \citep{benitezllambay2016fargo3d,benitezllambay2019fargo3dmf},
Idefix \citep{lesur2023idefix},
MPI-AMRVAC \citep{porth2014mpiamrvac,keppens2023mpiamrvac}, and the
SPH code Phantom \citep{price2018phantom}.

The third meaning is a multi-ion plasma: every ionic species
evolves its own velocity field and its own thermal pressure, the
usage established in space-weather modeling by BATS-R-US
\citep{toth2012batsrus} and GAMERA
\citep{zhang2019gamera}.

Throughout this paper, ``multi-fluid'' refers exclusively to neutral
and ion species: each carries its own velocity and thermal pressure
and is coupled through table-driven source terms $\mathcal{L}_c$.
Electrons are not an additional momentum fluid; in ionized
configurations they enter through a pressure-only quasi-neutral
electron closure. The formulation is hydrodynamic; photoionization,
recombination, and collisional exchange enter through the same
table-driven declaration.

Among the open-source astrophysical codes surveyed here, native
multi-fluid support is almost invariably of one form: a single thermally active gas coupled to
one or more pressureless dust species through a drag source term, as in
FARGO3D \citep{benitezllambay2016fargo3d,benitezllambay2019fargo3dmf},
Idefix \citep{lesur2023idefix}, MPI-AMRVAC
\citep{porth2014mpiamrvac,keppens2023mpiamrvac}, and the SPH code
Phantom \citep{price2018phantom}.

A second usage tracks several chemical species as co-moving fields
that share the single velocity and pressure of the bulk flow, as in
the multi-species facilities of FLASH \citep{fryxell2000flash} and
RAMSES \citep{teyssier2002ramses}, or the chemistry module
distributed with PLUTO \citep{mignone2012pluto}. Neither form gives
each species its own thermal pressure.

Charged multi-fluid systems, in which each ionic species carries its
own momentum and thermal pressure, are native degrees of freedom essentially
only in space-weather codes: BATS-R-US \citep{toth2012batsrus} and
GAMERA \citep{zhang2019gamera}; the multi-fluid equation set of the
latter is derived separately \citep{lyon2025gamera}.

The most direct peer of the present work is the multi-fluid dust module
of \citet{huangbai2022} --- the only dedicated multi-fluid module paper
in the Athena family --- which integrates the drag source implicitly
and ships its own suite of numerical tests; its fluids, however,
remain pressureless dust.

What is missing is a code in which several thermally active fluids,
neutral and charged alike, are native constituents of the same solver
and are declared rather than hard-coded. Xenia responds to this gap by
combining (i) multiple thermally active
fluids, (ii) a table-driven source coupling $\mathcal{L}_c$ declared
from Python, (iii) forward automatic differentiation of the symbolic
source-term IR, (iv) a quasi-neutral electron-pressure closure,
(v) operator-split radiation transport, (vi) a four-step Python user
interface, and (vii) single-node GPU portability from a single source.

The practical bottleneck is not another Riemann solver, but the cost of
turning a coupled multi-fluid model into a portable stiff-source
calculation. Xenia lowers that barrier by moving species definitions,
collision bands, chemistry, heating and cooling, and radiation channels
from kernel code to versioned Python tables. The same kernel evaluates
the resulting tapes, generates analytic forward-AD Jacobians, solves
the stiff sources cell by cell, and uses one Kokkos loop body for the
compiled Serial/OpenMP/CUDA builds.

Xenia is a ground-up rewrite of the private layer described above,
not a fork of an existing distribution. Three limitations of that
predecessor shaped the design: its chemical species were hard-wired
into C enumerations, so adding or removing a species meant editing
kernel code; its configuration lived in preprocessor macros and in
the line order of an INI-style input file, so the meaning of an
input could depend on its position; and its parallelism was bound
to MPI. Xenia replaces all three. Fluids and their couplings are
declared as data, in Python composition tables; no species name
and no problem-specific branch ever reaches the compiled kernels.
The user surface is the four-step API --- setup, step, fields,
close --- covering the whole life cycle of a run: declaring the
problem, advancing it in time, reading its fields back into
Python, and closing it down. The kernels are written once against
Kokkos and run on CPUs or on a single GPU from the same source.
The field layout is three-dimensional from the start;
one-dimensional problems are a degeneration of it, not a special
case. The design target is a single workstation or a single GPU
node, not a cluster allocation.

The architecture follows where the field is already going. The
2023--2026 generation of astrophysical code papers includes a prominent
one-source, many-backend portability trend: Idefix
\citep{lesur2023idefix}, AthenaK
\citep{stone2026athenak}, and KHARMA
\citep{prather2024kharma} are written against Kokkos, and gPLUTO
\citep{rossazza2026gpluto} against OpenACC. In each case
the physics is written once and the backend carries it to the
hardware at hand. What we take from these codes is the
portable-backend discipline, not yet another Riemann solver ---
the hyperbolic machinery of Xenia is standard. What none of the
distributions surveyed here provides, and what therefore has to be grown
in-house, is the multi-fluid part: the composition tables, the
source-term declarations, and the H/He/electron closure that a
given problem needs.

Table~\ref{tab:intro-code-matrix} condenses the comparison and the
backend trend. The table is not an exhaustive feature inventory of every
distribution; it records the capabilities relevant to the coupled-species
problem discussed here. For Xenia, magnetic physics and distributed
execution are deliberately marked as future capabilities: the claims of
this paper stop at the single-node HD release.

\begin{table*}[t]
\centering
\caption{Capability map of the representative code families discussed in
this introduction. Entries summarize the capability relevant to Paper~1
rather than every optional module in a distribution. ``Future'' marks the
Xenia series roadmap, not a claim of the present HD release.}
\label{tab:intro-code-matrix}
\fontsize{7.1}{8.6}\selectfont
\setlength{\tabcolsep}{2.5pt}
\begin{tabular}{p{0.130\textwidth}p{0.130\textwidth}p{0.070\textwidth}p{0.120\textwidth}p{0.075\textwidth}p{0.100\textwidth}p{0.120\textwidth}p{0.100\textwidth}}
\hline
Code family / representatives & Native species model & Thermal \(P\) per species & Source declaration & \(P_e\) closure & Radiation & MHD; distributed execution & Single-source portability \\
\hline
Reactive networks: FLASH, RAMSES, PLUTO + chemistry & Abundance fields advected with one bulk gas & No & Compiled problem/chemistry modules & --- & Code-specific modules & MHD in host codes; MPI/AMR & Varies; primarily conventional HPC \\
Gas + pressureless dust: FARGO3D, Idefix, MPI-AMRVAC, Phantom, Athena++ dust module & One gas plus dust velocities without dust pressure & Gas only & Module-specific drag setup & --- & Varies; not central to dust coupling & Host dependent; commonly MPI/AMR & Idefix uses Kokkos; others vary \\
Space-weather multi-ion: BATS-R-US, GAMERA & Multiple ion fluids & Yes & Compiled physics modules & Yes & Space-environment modules & MHD; large parallel frameworks & Conventional HPC \\
Portable Godunov frameworks: Athena / Athena++, AthenaK, gPLUTO, KHARMA & Primarily single-fluid gas/plasma; optional pressureless extensions & No general native set & Code modules & --- & Code-specific & HD/MHD or GRMHD; distributed support varies by generation & Central design; Kokkos or OpenACC \\
\textbf{Xenia Paper~1} & Neutral and ion thermally active fluids & \textbf{Yes} & Python tables + symbolic source IR & \textbf{Yes; pressure-only} & Radial Beer--Lambert scan & \textbf{Future}: MHD (Paper~2); MPI/AMR (Paper~3) & Serial / OpenMP / CUDA from one Kokkos loop body; single node \\
\hline
\end{tabular}
\end{table*}

Beyond filling the gap catalogued above, the present paper makes
three methodological contributions: (i) a table-driven multi-fluid
framework, in which a composition table and its symbolic source
tapes assemble the coupling terms $\mathcal{L}_c$ --- changing a
problem means changing a table, not a kernel --- and the
transonic, radiation-coupled escaping atmosphere of HD 209458b is
simply the initial table; (ii) a
pressure-only quasi-neutral electron closure, in which electrons
are not an independent momentum fluid: quasi-neutrality fixes the
electron density from the ion populations, the electron bulk
velocity is derived from those populations, only the electron
pressure $P_e$ is advanced, the
fluid-level effect of the ambipolar electric field enters every
ion species through the charge-weighted term $w_s P_e$ in its
Godunov flux, and the share of shock heating that goes to
electrons is an explicit option rather than a hidden default; and
(iii) an operator-split source treatment, in which every step runs
the non-local radiation scan first, then the hyperbolic flux
operator $L_f$, then the cell-local implicit solve for the stiff
coupling terms $\mathcal{L}_c$; forward automatic differentiation
over the same symbolic tapes generates an analytic, no-finite-difference
source Jacobian $J$; a single backward-Euler linearization per cell is solved
by a dense LU factorization, the radiation field enters the local
step as a frozen external source, and $J$ never couples across
cells --- precisely the locality that allows the source step to
run on a GPU.

\section{The Xenia Framework}
\label{sec:framework}

This section describes the Xenia software stack; the discretizations
themselves are deferred to the Numerical Methods section. The organizing principle
is that \emph{the physics is data}: which fluids exist, how they couple, and
which source terms act on them are declared in Python configuration tables,
while the compiled kernel contains no problem-specific code at all. Two
precedents shape the exposition. Athena++ devotes an entire section to
software structure before any equation appears
\citep{stone2020athenapp}, and the Space Weather Modeling Framework presents
its layering together with a catalog of interchangeable equation systems
\citep{toth2012batsrus}. Xenia follows the same genre: a layered framework
whose selling point is that exchanging a composition table exchanges the
problem being solved, with the kernel binary untouched.

\begin{figure}[t]
\includegraphics[width=0.98\textwidth]{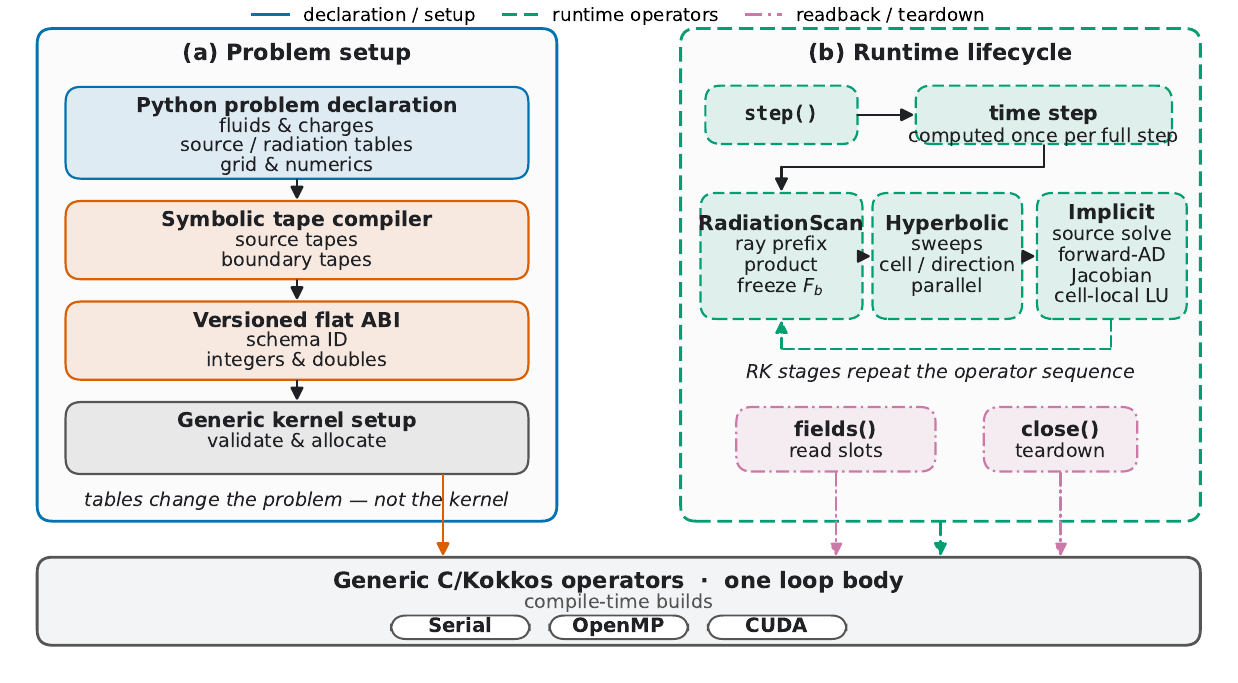}
\caption{The Xenia framework shown in setup and runtime views. (a) A Python
problem declaration---fluids, source and radiation tables, grid, and
numerical choices---is compiled into source and boundary tapes and packed
into the versioned flat ABI; generic kernel setup validates the schema and
allocates the run. Changing the problem therefore changes data crossing the
ABI, not the generic kernel. (b) One call to \texttt{step()} computes the
CFL time step once and then performs the RadiationScan, hyperbolic sweeps,
and implicit source solve; Runge--Kutta stages repeat this operator
sequence. \texttt{fields()} reads state slots back and \texttt{close()}
releases the run. Both panels use generic C/Kokkos operators with one loop
body; Serial, OpenMP, and CUDA denote separate compile-time backend builds,
with no runtime switching.}
\label{fig:layers}
\end{figure}

The stack has three layers (Figure~\ref{fig:layers}): a Python semantic layer
that owns every name and formula the user writes; a versioned, flat,
language-neutral application binary interface (ABI) that carries only
integers and doubles; and a C/Kokkos kernel that reads slots and integer
opcodes and nothing else. The remainder of this section walks the layers and
the contract between them: the four-step Python API
(Section~\ref{subsec:api}), the composition tables and source tapes that
constitute a problem (Section~\ref{subsec:tables}), the field layout and its
dimension-independence contract (Section~\ref{subsec:layout}), the build and
deployment path together with schema versioning
(Section~\ref{subsec:build}), and the software-engineering practice around
the kernel (Section~\ref{subsec:se}).

\subsection{The Three-Layer Stack}
\label{subsec:layers}

The semantic layer is a plain Python dictionary. It declares the fluid table
(each fluid carries a mass, an electric charge, and a species role), the
symbolic source expressions together with their named constants, the
radiation field as spectral bins with per-absorber cross sections and an
incident flux table, the unit system, the grid geometry and optional
stretching, the numerical choices by name (Riemann solver, reconstruction,
Runge--Kutta integrator), the boundary condition per face, and the initial
condition. The module \texttt{config.py} normalizes this dictionary into a
typed problem record and validates it before anything reaches compiled code.

Symbolic expressions never cross the layer boundary as text. The compiler in
\texttt{source\_ir.py} parses each \texttt{sources[]} expression with the
Python \texttt{ast} module and lowers it to a tape of integer opcodes and
operand slots; the opcode set 0--9 (load constant, load slot, add, subtract,
multiply, divide, power, negate, exponentiate, logarithm) is frozen. New
physics may append named quantities --- additional write targets, additional
boundary tapes --- but never renumbers an opcode, so a tape compiled against
one kernel revision remains readable by the next. Boundary assignments are
compiled by \texttt{bc\_ir.py} with the same opcode set under an assignment
semantic, and are applied outside the implicit solve.

The ABI layer is a single flat, versioned structure, \texttt{XmfConfig}:
22 32-bit integers followed by doubles, with no variable-length content and
no name strings. Its first field is the schema version, and the integer
enumerations it carries (geometry, backend, electron mode, Riemann solver,
reconstruction, integrator, stretching, shock-heating tier, per-face boundary
type) mirror, value for value, lookup tables on the Python side, so both ends
of the boundary validate the same integers. The byte layout is frozen by
rule: when the shock-heating tier was introduced it took over a previously
reserved integer slot, moving a name but never shifting a byte position.
Variable-length payloads --- the compiled source tapes and boundary tapes ---
ride beside the blob as three flat arrays (a per-tape header of integers, an
opcode/argument array, and a constant array). A single ctypes binding boundary is the only place the two languages meet.

The kernel consumes slots and opcodes exclusively. Its responsibilities are
partitioned among setup and teardown, mesh geometry and state packing,
direction-split Godunov sweeps, tape evaluation with forward-mode Jacobians,
the cell-local implicit source solve, the non-local radiation scan, compiled
boundary application, ABI and run handles, and Kokkos backend dispatch. An
ideal-MHD superset is also present in the source tree but is gated off by
default, reduces bit-wise to the zero-field limit when closed, and is reserved
for Paper~2. The public release documents the mapping from these
responsibilities to its source layout.

\subsection{The Four-Step Python API}
\label{subsec:api}

Every example, test, and application chain in this paper drives the kernel
through the same four verbs. \texttt{setup(cfg)} normalizes and validates the
user dictionary, compiles the source and boundary tapes, packs the flat blob,
and calls into the kernel, returning a run object; a schema mismatch or an
unsupported enumeration is refused here, at the boundary, rather than
surfacing later as a corrupted solve. \texttt{step()} performs one complete
time step under the kernel's CFL choice --- the operator-split chain of
radiation scan, hyperbolic update, and implicit source step that the
Numerical Methods section details --- and returns the new time.
\texttt{advance(tstop)} is a thin loop wrapper over \texttt{step()} for runs
that do not need per-step intervention, so scripting and interactive use read
identically. \texttt{fields()} hands back, per fluid, the cell coordinates,
number densities, velocities, and pressures as NumPy arrays, and
\texttt{close()} releases the run; at the C ABI layer the teardown symbol is
named \texttt{xmf\_finalize}, and that name appears in this paper only when
the ABI itself is the subject.

The four verbs are also where the framework's central design constraint is
enforced. The kernel is forbidden to branch on physical process names: there
is no \texttt{photoion} case in C, no hard-wired point-source gravity
$-GM/r^2$, and no recognition of species-name strings anywhere below the
semantic layer. A body force, a chemical rate, or a Coulomb exchange enters
the kernel only as a compiled tape whose write targets are slot offsets. The
consequence is the property this paper claims: changing the problem means
changing the table, not the code.

\subsection{Composition Tables and Source Tapes}
\label{subsec:tables}

Table~\ref{tab:tables} catalogs the nine configuration families retained in
this paper in the style of the component-by-model catalog of the Space Weather
Modeling Framework \citep{toth2012batsrus}. Each row is a composition table
(which fluids, with which charges and roles) crossed with the source tapes and
radiation declaration that couple or drive them. A family can instantiate
several numerical variants---for example solver choices in \texttt{hd\_eval}
or coupling tiers in the cloud and jet families---so the table records the
invariant structure rather than every numerical knob. All nine families run
against the same kernel binary and four-step API; changing family or tier
changes data, not compiled problem code.

The first six rows isolate neutral hydrodynamics and neutral multi-fluid
coupling. The empty-source rows implement the hydrodynamic limit by
declaration, \texttt{sources: []}, not by tuning rate coefficients to zero.
The final three rows escalate from radiation-coupled hydrogen hydrodynamics to
four ionized species with two absorbers and the pressure-only electron
closure. Table~\ref{tab:tables} intentionally excludes solver, reconstruction,
time-integrator, and backend settings: those belong with the corresponding
Code Validation records. Every structural entry is read from a committed
configuration and an archived v0.1 report; it is not a simulation result.
\begin{table*}
\caption{Composition/configuration families used throughout this paper. A
family may instantiate several numerical variants. $N$ is the dimension of the
implicit source system ($5$ stored slots per fluid, plus one when the
pressure-only electron closure is active); in that closure $P_e$ is the single
transported electron scalar, not a momentum fluid. Beam counts refer to the
radiation spectral table, and absorbers are the species over which the scan
accumulates optical depth. Solver, reconstruction, time-integrator, and backend
settings are deferred to the validation records.}
\label{tab:tables}
\small
\setlength{\tabcolsep}{4pt}
\begin{tabular}{@{}p{0.14\textwidth}p{0.14\textwidth}p{0.24\textwidth}cp{0.13\textwidth}p{0.15\textwidth}@{}}
\hline
Configuration family & Fluids & Source tapes & $N$ & Radiation & Use in paper \\
\hline
\texttt{hd\_eval} & 1 neutral (HI) & empty (\texttt{sources: []}) & 5 & off & HD solver and geometry tests \\
\texttt{dustybox\_1d} & A, B neutral & 1 momentum-drag tape & 10 & off & drag benchmark \\
\texttt{mf\_cloud\_2d} & wind, cloud neutral & empty or 2 momentum-drag tapes (coupling tier) & 10 & off & cloud--shock \\
\texttt{mf\_jet\_2d} & light, heavy neutral & empty or 2 momentum-drag tapes (off/on) & 10 & off & jet coupling \\
\texttt{mf\_blast\_3d} & light, heavy neutral & empty & 10 & off & 3D momenta \\
\texttt{cyl\_2d} & gas neutral & empty & 5 & off & wake capability \\
\texttt{rad\_hd\_1d} & HI, HII & 10: photoionization, recombination, drag, charge-exchange, heating, gravity/tide & 10 & 53 beams, HI absorber & radiation HD \\
\texttt{rad\_h\_pe\_1d} & HI, HII; $P_e$ & 16: adds electron heating and Coulomb/collisional exchange & 11 & 53 beams, HI absorber & electron pressure \\
\texttt{rad\_hhe\_pe\_1d} & HI, HEI, HII, HEII; $P_e$ & 26: chemistry, drag, charge/thermal exchange, gravity & 21 & 53 beams, HI and HEI absorbers & four-fluid chain \\
\hline
\end{tabular}
\end{table*}

The implicit-system dimension quoted in the table is a contract, not an
observation: the run object reports $N = 5\,n_{\mathrm{fluid}} + 1$ when the
electron closure is active and $5\,n_{\mathrm{fluid}}$ otherwise, because the
electron pressure occupies its own state variable and never enters the five
fluid slots. Spectral resolution behaves the same way: the 53-bin radiation
table, its per-absorber cross sections, and the incident flux are data loaded
at setup, so a configuration may collapse to a single-bin fallback or expand
to a finer table without any change below the semantic layer.

\subsection{Field Layout and the 3D-Ready Contract}
\label{subsec:layout}

The kernel stores the fluid state in one flat array indexed
$[\mathrm{fluid}][\mathrm{slot}][k][j][i]$, with cells ordered $(k,j,i)$ and
the five slots per fluid fixed as $(n, v_x, v_y, v_z, P)$. The pressure-only
electron variable, when active, is appended as a sixth state of the problem
rather than as a slot of any fluid, preserving the stride of the fluid block.
A one-dimensional run is the degenerate case $(1,1,n_x)$: the array shape
changes, but every loop in the kernel keeps three indices and every sweep
iterates the same triple-nested structure. Dimensionality is therefore a
property of the configuration, not of the code path, which is what lets a
single kernel serve one-, two-, and three-dimensional problems from one
source; the geometry enumeration (Cartesian, spherical, cylindrical) enters
only through face areas and cell volumes, whose construction belongs to the
hyperbolic operator of the Numerical Methods section. Ghost zones are carried per
face, six of them, and the compiled boundary tapes fill them.

\subsection{Build, Deployment, and Schema Versioning}
\label{subsec:build}

The kernel is built with CMake (C99 for the orchestration and mesh layers,
C++17 for the sweep and source layers) and packaged as a wheel through
scikit-build-core; Kokkos enters as a vendored third-party submodule at
version 4.7 and provides the parallel constructs for three compile-time
backends --- Serial, OpenMP, and single-device CUDA. The Python run
configuration declares the intended backend, and setup validates that name
against the backend enumeration compiled into the library. One compiled
library exposes exactly one backend; there is no runtime backend switching.
Because the loop bodies are shared, Xenia offers single-source deployment
across these target builds. This is a functional-portability statement, not
a performance-portability claim; the archived evidence and its limitations
are treated in the Backend Equivalence discussion of the Code Validation section.

Schema versioning closes the loop between the Python and C ends of the ABI.
The semantic layer publishes a single integer schema version; the header
publishes the same integer as \texttt{XMF\_SCHEMA}; and the flat blob carries
the version as its first field. At setup the kernel validates the field and
refuses any configuration that does not match, so a stale Python front end
cannot silently feed a newer kernel, or the reverse. The frozen byte-position
rule of Section~\ref{subsec:layers} is the companion discipline on the C
side: fields may be renamed and reserved slots may be claimed, but no field
may be inserted where it would shift the layout an older build still trusts.

Deployment has exactly one supported path. The repository ships a sync
script that builds the kernel and copies the compiled library together with
the Python sources into the live import tree; because the native loader reads
the library from beside its own module, any other copy --- a build directory,
an installed wheel from an earlier day --- is a different binary, and
reproduction instructions in this paper point at the sync script exclusively.
Running the regression suite against an unsynced tree is, by construction, a
measurement of the wrong artifact.

\subsection{Software Engineering Practice}
\label{subsec:se}

The v0.1 release suite records 394 passing pytest tests across forty test
files, spanning the layers of the validation pyramid of the Code Validation
section: unit contracts on the tape compiler and the automatic-differentiation
Jacobians against analytic derivatives, conservation regressions on closed
source systems, contract tests on the radiation scan and the electron
closure, and backend-equivalence checks that pin the Serial result and bound
the OpenMP and CUDA deviations.

Development proceeds in slices: each change carries a card that states its
scope, its pre-declared acceptance criteria, and its evidence, is frozen at
the moment work begins, and is retained after closure, so that the history of
the kernel reads as a sequence of audited increments. This practice follows
the observation, made in the Space Weather Modeling Framework paper, that a
code paper should report its engineering discipline --- coding standards and
nightly regression testing there \citep{toth2012batsrus} --- and not only its
numerics.

\section{Numerical Methods}
\label{sec:methods}

The framework described in Section~\ref{sec:framework} advances the state of
a run by one complete time step per \texttt{step()} call. This section
documents the continuous model, its notation, and the numerical operators
that a step applies.

\paragraph{Continuous model and notation.}
Let $s$ index a thermally active neutral or ion species. Its HD state is
\begin{equation}
U_s=\left(\rho_s,\rho_s\mathbf v_s,E_s\right),\qquad
E_s=\frac{P_s}{\gamma-1}+\frac{1}{2}\rho_s|\mathbf v_s|^2,
\label{eq:species-state}
\end{equation}
with $\rho_s=m_s n_s$ its mass density and $E_s$ its total energy density.
The stacked multi-fluid state $U$ collects these species vectors; when the
electron closure is active, one scalar electron pressure $P_e$ is appended.
Continuously, the solved model is
\begin{equation}
\frac{\partial U}{\partial t}+\nabla\!\cdot\!\mathbf F(U)
=\mathcal{L}_c(U,r,F_b).
\label{eq:continuous-model}
\end{equation}
Here $\mathbf F$ denotes the generic multi-fluid HD flux, deliberately not
written componentwise at this entry: its discrete Riemann construction is the
subject of Section~\ref{subsec:hyperbolic}. In ionized configurations the ion
Riemann pressure is the effective pressure \(P_{\mathrm{eff},s}=P_s+w_s\,P_e\)
specified in Section~\ref{subsec:electron}, with \(w_s=q_s\,n_s/n_e\) for
charged species and \(w_s=0\) for neutrals. Electrons are not an additional
momentum fluid: no electron mass or momentum is advanced, and \(P_e\) is the
one transported electron scalar. The symbol \(F_b\) denotes transmitted beam
fluxes computed by the ray scan; after that scan it is frozen for the step
and is not differentiated by the source Jacobian. The solver realizes
configured coordinate geometry through mesh face areas and cell volumes. Its
equations and mesh arrays are expressed in the configured unit system, while
the radiation scan converts ray segment lengths to physical units when it
forms optical depths (Section~\ref{subsec:radiation-scan}).

For operator notation, the hyperbolic operator \(L_f\) below represents the
discrete negative of the flux divergence, and the implementation splits the
right-hand side of the evolution into two operators acting on $U$ --- five
conservative slots per fluid (mass density, three momentum densities, and
energy density), plus one trailing electron-pressure slot when the closure is
active:
\begin{equation}
\frac{\partial U}{\partial t} = L_f(U) + \mathcal{L}_c\left(U;\, r,\, F_b\right),
\label{eq:split}
\end{equation}
where $L_f$ is the hyperbolic flux operator, $\mathcal{L}_c$ collects all
table-driven source terms, $r$ is the grid coordinate, and $F_b$ are the
radiation beam fluxes. One step executes three sub-stages in a fixed order.
First, the radiation scan (Section~\ref{subsec:radiation-scan}) recomputes
the transmitted beam fluxes $F_b$ from the current absorber densities. The
scan is the only non-local operator: rays are independent, but transport is
serial along each ray. Second, the hyperbolic operator $L_f$
(Section~\ref{subsec:hyperbolic}) advances every fluid by explicit,
cell- and direction-parallel sweeps. Third, the stiff source operator
(Sections~\ref{subsec:source-ir} and~\ref{subsec:implicit}) applies the
chemistry, photochemistry, drag, and heating terms cell-locally and
implicitly. After the scan, $F_b$ is frozen for the remainder of the step:
the source tapes read it as a known external field, and it is never
differentiated. The explicit Runge--Kutta integrator wraps the whole-step
operator: every stage repeats the same three-sub-stage sequence around one
evaluation of $L_f$, and $\Delta t$ is computed once per full step from $U^n$
and shared by every stage. The source operator and its Jacobian are
cell-local and contain no inter-cell coupling; the scan is the only part of
the step that traverses cells along a ray. Thus each sub-stage maps onto a
single parallel pattern --- ray-parallel with serial transport within each
ray, cell- and direction-parallel, and strictly cell-local, respectively. This
is the precondition that lets the whole step run on an accelerator without
inter-cell communication (Section~\ref{subsec:kokkos}).

\begin{figure}[t]
\centering
\includegraphics[width=0.98\textwidth]{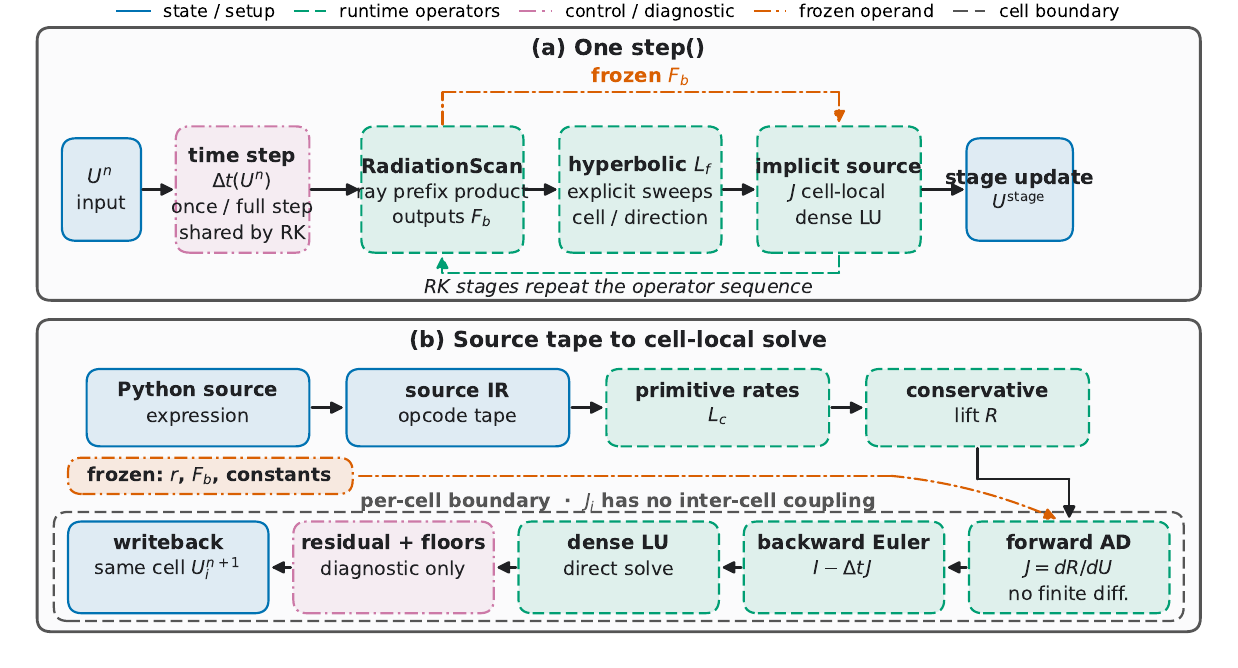}
\caption{Implementation structure of one Xenia step and of the table-driven
source solve. (a) A step first recomputes the transmitted beam flux
$F_b$ with the radiation scan, then applies the hyperbolic operator $L_f$,
and finally solves the stiff source terms implicitly; $F_b$ is frozen after
the scan. The CFL time step $\Delta t$ is computed once from $U^n$ for the
full step and shared by all Runge--Kutta stages, which repeat the same
operator sequence. (b) Python source expressions are compiled to
source-IR opcode tapes; tape evaluation gives primitive rates $L_c$, the
conservative lift gives $R$, and the same tapes provide the forward-AD
Jacobian $J=dR/dU$ without finite differences. Each cell then solves one
backward-Euler linear system by dense LU, reports the residual diagnostically,
applies floors, and writes the update back in that cell. The source Jacobian
and writeback have no inter-cell coupling. The diagram summarizes the
implemented structure; it introduces no new algorithm and makes no
performance claim.}
\label{fig:methods-operators}
\end{figure}

Figure~\ref{fig:methods-operators} summarizes the same fixed split and the
source-to-solve pipeline.

\subsection{The Hyperbolic Operator}
\label{subsec:hyperbolic}

\paragraph{Area-weighted finite-volume form.}
The flux operator $L_f$ is a direction-split Godunov-type finite-volume
scheme \citep{leveque2002book,toro2009book} in the tradition of the PLUTO
and Athena families \citep{mignone2012pluto,stone2020athenapp}. One sweep per
active direction updates the conservative state with area-weighted numerical
fluxes,
\begin{equation}
U_i \leftarrow U_i - \frac{\Delta t}{\Delta V_i}
\left(A_{i+1/2}\, F_{i+1/2} - A_{i-1/2}\, F_{i-1/2}\right),
\label{eq:fv-update}
\end{equation}
where $A$ is the face area, $\Delta V$ the cell volume, and $F$ the numerical
flux in the sweep direction. Each sweep consists of two kernels: a read-only
face kernel that evaluates $F$ at every interior interface of that direction
into a face-flux buffer, followed by a cell kernel that applies each cell's
own two buffered fluxes. Because face evaluation writes no state, both
kernels are parallel across all faces, respectively cells, of the plane
perpendicular to the sweep.

Geometry enters Equation~\ref{eq:fv-update} exclusively through the
precomputed face areas and cell volumes that the mesh layer fills at setup.
Cartesian coordinates carry $A = 1$ and $\Delta V = \Delta x$; in spherical
coordinates the radial faces carry $r^2$ and the polar faces carry
$|\sin\theta|$; in cylindrical coordinates the radial faces carry $|R|$.
There is no geometry-specific branch anywhere in the solver: the same face
and cell kernels run in all three coordinate systems, and a uniform Cartesian
grid reduces bit-for-bit to the classic $\Delta t/\Delta x$ update.

\paragraph{Interface fluxes and reconstruction.}
Face states feed an approximate Riemann solver chosen at setup among three
implemented options: HLL \citep{harten1983hll}, HLLC \citep{toro1994hllc},
and a total-variation-diminishing Lax--Friedrichs variant
\citep{harten1983tvd,toth1996fct}. All three carry the full interface state
--- the direction-normal velocity plus both tangential velocities --- and
tangential momentum rides the mass flux. Cell states are reconstructed to
the interfaces by one of three selectable tiers: piecewise constant
(first order), piecewise linear with the van Leer slope limiter
\citep{vanleer1977}, or third-order weighted essentially non-oscillatory
reconstruction \citep{jiang1996weno,toro2009book}. The reconstruction acts
componentwise on the primitive variables; no characteristic decomposition is
performed, and we state this as the deliberate implementation simplification
it is rather than as a characteristic-space scheme. Solver and reconstruction
selections are validated configuration enumerations: unsupported names are
refused at the setup boundary (Section~\ref{subsec:api}), so a run never
silently degrades to a different scheme.

\paragraph{Time integration.}
The explicit integrator is a strong-stability-preserving Runge--Kutta scheme
of order one, two, or three --- forward Euler, Heun's method, and the
third-order scheme of Shu and Osher --- with the standard SSP coefficients
\citep{gottlieb2001ssp}. The integrator wraps the whole-step flux operator:
one stage evaluation performs all active-direction sweeps. A fourth-order
entry exists in the configuration vocabulary but is refused by the setup
validation; enumerated names and implemented capabilities never diverge
silently.

\paragraph{Multi-fluid sweeps and the CFL condition.}
Each fluid is swept independently over its own five conservative slots: the
hyperbolic fluxes couple no species to any other. All inter-species coupling
--- chemistry, photochemistry, drag, heating --- lives in the source
operators (Sections~\ref{subsec:source-ir} and~\ref{subsec:implicit}), which
keeps the hyperbolic stage a set of independent Euler solves sharing one
geometry. The time step follows the usual signal-speed criterion, evaluated
once per full step from $U^n$:
\begin{equation}
\Delta t = \frac{\mathrm{cfl}}{\max_{\mathrm{cell}}
\sum_{d} s_{d}/\Delta x_{d}},
\label{eq:cfl}
\end{equation}
where the sum runs over the active directions only, and the per-cell,
per-direction signal speed $s_d$ is the maximum over the fluids of
$|v_d| + c_s$ at that cell. When the electron closure is active, the ion
sound speed is computed from the effective pressure of
Section~\ref{subsec:electron}, and electron-pressure transport
contributes its own signal speed $|u_{e,d}| + c_e$ to the same per-cell
maximum. A user-specified $\Delta t_{\max}$, when positive, caps the result
from above; cell widths come from the per-cell mesh arrays, never from a
scalar spacing, so stretched grids are handled by the same formula.

\subsection{Table-Driven Source Terms and the Source IR}
\label{subsec:source-ir}

\paragraph{From formula strings to integer tapes.}
Every source term in Xenia is data, not code --- the zero-specialization rule
of the three-layer stack (Section~\ref{subsec:layers}). Users declare rate
laws as Python formula strings in composition tables: Arrhenius rates, power
laws, photo-rates of the form $\sigma F / h\nu$, Coulomb drag, gravitational
momentum and energy exchange. At setup, the semantic layer parses each string
with the Python \texttt{ast} module and compiles it into a tape of integer
opcodes drawn from the frozen ten-instruction set (opcodes 0--9: load
constant, load slot, add, subtract, multiply, divide, power, negate,
exponential, logarithm). Operands are typed by kind --- state slot, grid
coordinate $r$, beam flux $F_b$, or an embedded numeric constant --- and the
Python-side opcode table is kept in lockstep with the C-side evaluator. At
run time the kernel evaluates tapes and nothing else: it parses no text,
generates no code, and instantiates no per-network functions, so swapping one
chemistry network for another never touches compiled code.

\paragraph{The lift into conservative rates.}
Per fluid, the tape rates arrive in primitive form,
$L_c = (\dot n, \dot v_x, \dot v_y, \dot v_z, \dot P)$, over the primitive
state $w = (n, v_x, v_y, v_z, P)$ with $n = \rho/m$ the number density of the
fluid of mass $m$. A fixed lift $T$ converts them into conservative rates
$R = T(w, L_c)$: the mass row is $\dot\rho = m\,\dot n$, the momentum rows
combine mass loading with direct acceleration,
$\dot{(\rho \mathbf v)} = m\,\dot n\,\mathbf v + \rho\,\dot{\mathbf v}$, and
the energy row reads
\begin{equation}
\dot E = \frac{\dot P}{\gamma - 1}
+ \frac{1}{2}\,|\mathbf v|^2\,\dot\rho
+ \rho\,\mathbf v\cdot\dot{\mathbf v}.
\label{eq:lift}
\end{equation}
The same fixed lift is used both to add evaluated tape rates to the state
and in Jacobian assembly below; there is no separate hand-coded conservative
source path.

\paragraph{The differentiability contract.}
The implicit source step of Section~\ref{subsec:implicit} requires analytic
derivatives of the rates with respect to the state, which fixes the function
class a tape may express: only differentiable rate laws are accepted. This
restriction matches the default assumption of the major chemistry toolchains
(Cantera, KROME, pyJac), whose exported networks are analytic in the state
almost everywhere --- Arrhenius factors, power laws, $\sigma F/h\nu$
photo-rates, and Coulomb drag are all smooth. Piecewise-defined rates with
hard switches are deliberately outside the contract: a Newton-type solve
needs a consistent Jacobian rather than an averaged subgradient at the kink,
so thresholds belong in the table design --- smoothed, or resolved into
separate continuous regimes --- not in the tape evaluator.

\paragraph{Forward-mode automatic differentiation as a stack machine.}
Because the networks are data, their derivatives must be data-derived:
maintaining hand-written analytic Jacobians for arbitrary user-supplied rate
expressions is unmanageable, and forward-mode automatic differentiation is
the natural consequence of the table-driven design \citep{baydin2018}.
The derivatives are generated from the same opcode tapes that evaluate the
rates; there is no independently coded analytical source model.
Xenia's implementation is a hand-written forward-mode dual-number stack
machine inside the kernel: every stack entry carries a pair $\{v, \dot v\}$,
and each of the ten opcodes propagates the value and its derivative by the
matching chain rule, with state-independent operands entering at
$\dot v = 0$. The machine is written once as \texttt{KOKKOS\_INLINE}
functions, so the host and device paths execute identical arithmetic ---
there is no separate device-side differentiation code. Seeding is explicit:
an evaluation names one unknown slot whose derivative starts at unity while
all others start at zero; an evaluation that names no slot leaves every
derivative at zero and reproduces the non-differentiating value path
bit-for-bit. Applying rates and differentiating them therefore never
constitute two numerically distinct code paths. Embedding hand-written
forward mode directly in a finite-volume kernel follows the same line as
recent work on automatic differentiation for Cauchy--Kovalevskaya procedures
\citep{babbar2026}; the treatment of implicit functions follows
\citet{margossian2021}.

\paragraph{Jacobian assembly.}
The unknowns of the source system are exactly the stored conservative slots
of each fluid --- five per fluid --- plus, when the electron closure is
active, the one trailing electron-pressure unknown, so the system size is
$N = 5\,n_{\mathrm{fluid}}\,(+1)$. The per-cell Jacobian is assembled by
seeding each unknown slot in turn and collecting the tape derivatives into a
dense $N \times N$ matrix $J = dR/dU$: the same tapes supply the
primitive-rate derivatives, and the lift derivatives complete the conversion
to conservative unknowns. The grid coordinate $r$, the beam
fluxes $F_b$, and the embedded constants are never seeded: frozen column
masks in the tapes skip every column those operands would otherwise touch,
so these dependencies stay frozen inside the step. The full derivative is
the four-segment chain
\begin{equation}
\frac{dR}{dU} = \left(S_1 S_2 + S_4\right) S_3,
\label{eq:chain}
\end{equation}
in which $S_1 = \partial T/\partial L_c$ and $S_3 = \partial w/\partial U$
are block-diagonal lifts between primitive and conservative variables,
$S_2 = \partial L_c/\partial w$ is the forward-mode pass over the loaded
tapes, and $S_4 = \partial T/\partial w$ at frozen $L_c$ captures the state
dependence of the lift itself. $S_4$ is mandatory: omitting it would
differentiate the rates but not their conversion into conservative form. All
fluid blocks are $5 \times 5$; the electron-pressure row is trivial, since
$P_e$ is itself the unknown and the lift acts as the identity, so its
gathered tape rate enters the Jacobian as-is.

\subsection{The Implicit Source Step}
\label{subsec:implicit}

\paragraph{One linearized backward-Euler solve per cell.}
The stiff source substep integrates every cell with a single-linearization
backward-Euler step of Rosenbrock-1 form:
\begin{equation}
\left(I - \Delta t\, J\right) dU = \Delta t\, R(U^*),
\qquad
U^{n+1} = U^* + dU,
\label{eq:implicit}
\end{equation}
where $U^*$ is the state entering the source substep, $R$ the lifted tape
rates of Section~\ref{subsec:source-ir}, and $J = dR/dU$ the chain-rule
Jacobian of Equation~\ref{eq:chain} evaluated once at $U^*$. Each cell
receives exactly one dense $N \times N$ LU factorization with partial
pivoting and one direct triangular solve per step. There is no nonlinear
iteration and no Jacobian-free Newton--Krylov machinery, by design: at the
small system sizes of this paper --- $N = 11$ for the two-fluid
configurations with the electron-pressure unknown, $N = 21$ for the
four-fluid ones --- forming the dense Jacobian and factorizing it directly is
the intended default. The Jacobian is analytic and finite-difference-free, and
the linear system is solved by direct dense LU factorization followed by
triangular solves. Floating-point roundoff and pivoting still enter; the claim
is source consistency and a direct linear solve, not mathematical exactness in
floating-point arithmetic. This gives a predictable per-cell cost.
Iterative alternatives pay off only around $N \gtrsim 100$, where an
approximate Jacobian is needed as a preconditioner in any case.  The
semi-implicit coupling of tabulated multi-component rates follows the approach of
\citet{garciamunoz2007}, and the cell-local linearized form belongs to the
same family as the point-implicit source treatment of BATS-R-US
\citep[see][\S4.4]{toth2012batsrus}.

\paragraph{Residual reporting and floors.}
Because one linearization is the contract, the step reports its own error
instead of iterating on it: a relative maximum residual over the cells,
\texttt{source\_resid}, is computed at the end of the source substep and
returned to the caller. It is diagnostic only --- it never triggers a
re-solve inside the step; the behavior of the single-linearization step under
far-from-equilibrium initial data is documented in the Code Validation
section. After the solve, mass densities and pressures are floored at
$10^{-30}$ in code units before write-back, the same floor the hyperbolic
primitive recovery applies, so a stiff step can never inject a negative
density or pressure into the next sweep.

\paragraph{Cell locality.}
A tape reads only its own cell's slots, and every write lands in that same
cell: the implicit source step is strictly cell-local. It therefore rides a
single Kokkos parallel loop over the interior cells, with the residual
diagnostic as one parallel reduction --- one formula, identical on Serial,
OpenMP, and CUDA builds. Together with the frozen external field $F_b$ of
Equation~\ref{eq:split}, no cell requires another cell's data during the
source substep, which is what makes the stiff part of the step trivially
accelerator-friendly.

\subsection{Quasi-Neutral Electron Pressure Closure}
\label{subsec:electron}

\paragraph{A closure, not a fifth fluid.}
Ionized configurations need an electron pressure, but not a full electron
momentum fluid. Xenia follows the multi-fluid ionosphere tradition, in which
the electron population enters through a pressure-only quasi-neutral closure
rather than a momentum equation \citep{schunk2009ionospheres} --- the same
model class used for planetary ionospheres and escaping atmospheres
\citep{ma2019,xing2023,xing2024}. Quasi-neutrality fixes the electron number
density from the ions,
\begin{equation}
n_e = \sum_{s \in \mathrm{ions}} q_s\, n_s,
\label{eq:quasineutral}
\end{equation}
with $q_s$ the charge number of species $s$, and the electron bulk velocity
is the charge-weighted ion velocity,
\begin{equation}
\mathbf u_e = \frac{1}{n_e} \sum_{s \in \mathrm{ions}}
q_s\, n_s\, \mathbf u_s.
\label{eq:ue}
\end{equation}
No electron mass or momentum is ever advanced. The single scalar $P_e$ is
appended to the state as the trailing slot
(Section~\ref{subsec:source-ir}), is transported by the electron-pressure
sweep and is
exchanged with the ions through source terms --- for instance Coulomb
relaxation --- like any other tape-driven quantity.

\paragraph{Effective pressure in the ion Godunov fluxes.}
The electrons act on each ion species inside its Riemann problems through an
effective pressure,
\begin{equation}
P_{\mathrm{eff},s} = P_s + w_s\, P_e,
\qquad
w_s = \frac{q_s\, n_s}{n_e},
\label{eq:peff}
\end{equation}
which replaces $P_s$ in the ion sweeps of all three directions: the electron
pressure force is split among the ions by charge share. Neutral species
carry $w_s = 0$ and hence $P_{\mathrm{eff},s} = P_s$; cells with degenerate
$n_e \le 0$ fall back to $w_s = 0$, $\mathbf u_e = 0$, and
$P_{\mathrm{eff},s} = P_s$, so the closure never amplifies a charge-free
region. The ion sound speed of the CFL condition
(Equation~\ref{eq:cfl}) reads $P_{\mathrm{eff},s}$, which keeps the
hyperbolic stage aware of the electron pressure support.

\paragraph{Two-pass integration and the advection of $P_e$.}
Because the Riemann solver sees $P_{\mathrm{eff}}$ while each species'
thermodynamics needs $P_s$, a single sweep cannot update both consistently.
Every direction sweep with the closure active therefore consists of two
integration passes. The first pass treats $(\rho_s, \mathbf v_s,
P_{\mathrm{eff},s})$ as the Riemann state and keeps the updated mass density
and momenta; the second pass treats $(\rho_s\, P_s/P_{\mathrm{eff},s},
\mathbf v_s, P_s)$ and keeps only the updated species pressure --- the ion
internal energy never rides $P_e$. Afterwards, $P_e$ itself is advanced by
one additional pseudo-fluid sweep built from $n_e$, the charge-weighted
velocity $\mathbf u_e$ of Equation~\ref{eq:ue}, and $P_e$, passing through
the same Godunov machinery and keeping only its pressure increment. That
sweep transports only the scalar $P_e$ and introduces no electron mass or
momentum equation. This
combination --- splitting the electron pressure by charge share into
per-species Godunov fluxes, integrating in two passes, and advecting $P_e$
with a dedicated pseudo-fluid sweep --- is rarely described in the public
method literature; the closest published analogue we are aware of is the
GAMERA code, which likewise distributes the electron pressure among the ions
by charge share \citep{zhang2019gamera,lyon2025gamera}.

\paragraph{Shock-heating partition: options, not truth.}
Collisionless shocks partition their dissipated energy between electrons and
ions through kinetic-scale physics that no fluid model resolves, and the
post-shock $T_e/T_i$ is not prescribed by the underlying physics. Xenia therefore exposes the partition as
three explicit, switchable tiers applied to the non-adiabatic ---
entropy-producing --- part of the interface pressure jump, following the
convention of \citet{toth2012batsrus}: ions-only (electron share
$f_s = 0$), by-charge ($f_s = w_s$), and electron ($f_s = 1$). The share
removed from an ion's pressure increment is deposited onto $P_e$ after that
direction's pseudo-fluid sweep, so the sweep itself stays tier-blind and the
reversible compression work of $P_e$ is untouched by the partition; deposits
occur only at cells whose ion density is above the floor. We state the
scope of these tiers explicitly: none of the three is physical truth. They
are documented model options that bound the electron-heating uncertainty,
and results obtained with them must be reported with the tier named.

\subsection{The Radiation Scan Operator}
\label{subsec:radiation-scan}

\paragraph{Rays and beam tables.}
Photoionization and photoheating rates require the local radiation field,
which the scan operator computes as a formal solution along prescribed rays.
In the mode used throughout this paper, a ray is an ordered list of grid
cells together with its segment lengths $\Delta s$ --- radial ray bundles for
the configurations considered here. Rays are mutually independent, so the
scan parallelizes over rays on the device while each ray's transport is
evaluated serially along its cells. The beam properties enter as tables
loaded at setup: per-beam photon energies $h\nu$ and incident fluxes at the
ray entrance, plus absorption cross sections $\sigma_{a,b}$ for every
absorber species $a$ and beam $b$, organized as a multi-absorber beam table
with an absorber index map, so any number of species can share one beam set.

\paragraph{Beer--Lambert prefix product and photon conservation.}
Along each ray the transmitted flux obeys Beer--Lambert attenuation,
implemented as a prefix product marching from the illuminated end of the
ray: cell $i$ receives the flux transmitted by its upstream neighbor and
attenuates it by its own optical depth,
\begin{equation}
F_{i,b} = F_{i+1,b}\, e^{-\Delta\tau_{i,b}},
\qquad
\Delta\tau_{i,b} = \sum_a \sigma_{a,b}\, n_{a,i}\, \Delta s_i,
\label{eq:prefix}
\end{equation}
where $\Delta s_i$ is the segment length of cell $i$ along the ray in
centimeters --- the grid length times the configuration's unit length --- and
$n_{a,i}$ is the number density of absorber $a$, obtained from its stored
mass density and per-species mass. The index convention is that cell index $i$
increases outward and the illuminated boundary is the high-index end:
$F_{i+1,b}$ is the flux transmitted out of upstream cell $i+1$ and therefore
entering cell $i$, while $F_{i,b}$ is the flux after attenuation in cell $i$.
The flux absorbed in each cell is
\begin{equation}
\Delta F_{i,b} = F^{\mathrm{in}}_{i,b}
\left(1 - e^{-\Delta\tau_{i,b}}\right),
\label{eq:thickcell}
\end{equation}
with $F^{\mathrm{in}}_{i,b}$ the flux entering the cell. The exponential form
of Equation~\ref{eq:thickcell}, rather than a linear truncation in
$\Delta\tau$, conserves photon number in optically thick cells: for
$\Delta\tau \gg 1$ the absorption saturates at the incoming flux, so
no cell can destroy more photons than arrive. The per-beam absorbed fluxes
are what the photo-rate tapes of Section~\ref{subsec:source-ir} read as
$F_b$.

\paragraph{A frozen external field, and the scope of the claim.}
After the scan, $F_b$ is a known external field for the remainder of the
step: the source tapes read it, and the Jacobian never differentiates it
(Section~\ref{subsec:source-ir}). No operator feeds the hydrodynamic state
back into the radiation field within the step, so the scheme is an
instantaneous formal solution, justified when the light-crossing time of the
domain is much shorter than the hydrodynamic step, $c\,\Delta t \gg R$. This
sentence also fixes the scope of what this paper claims: a prefix product of
Beer--Lambert attenuation along radial ray bundles, with photon-conserving
thick-cell absorption --- not a general radiation solver. Time-dependent
radiation transport --- ray-based transport schemes, flux-limited diffusion,
or moment closures --- is outside this paper and is deferred to the second
paper of this series.

\subsection{Kokkos Parallelization}
\label{subsec:kokkos}

\paragraph{One loop body, three backends.}
Every kernel described above --- face fluxes and cell updates, SSP staging,
the implicit source loop and its residual reduction, the ray scan, and the
boundary application --- is written once against the Kokkos interface
\citep{edwards2014kokkos} and compiles for Serial, OpenMP (host-parallel),
and single-device CUDA targets.
The single-source portable implementation strategy follows the line taken by AthenaK
\citep{stone2026athenak}: one source of truth for the arithmetic, builds per
target, and no duplicated numerics between host and device code --- in
Xenia's case down to the differentiation machine of
Section~\ref{subsec:source-ir}, which is a single
\texttt{KOKKOS\_INLINE} implementation shared by both.

\paragraph{Compile-time backend selection.}
Backend selection is a compile-time property: one compiled library, one
backend, no runtime switching. The execution space of every kernel is the
default execution space of the build --- the CUDA device in a device build,
the OpenMP host in a host-parallel build --- and every owning view lives in
the matching default memory space. The backend declared in the run
configuration is validated against its enumeration at setup and travels in
the configuration blob as an integer, but the kernel never consults that
field at runtime: the integer is a declaration-consistency record --- the
validated name that accompanies the run --- and not a switch that could
select a different backend inside a compiled library.

\paragraph{The device-resident trio.}
Three pieces of machinery keep the hot data on the device across a whole
step, so that no sub-stage forces a full-field round trip. First, the
boundary-condition state machine: boundary definitions are dispatched at
setup, applied on the device, and maintained lazily, filling the ghost skin
in place while an internal pin mask re-pins frozen interior cells, so
applying boundary conditions never copies the full field to the host and
back. Second, the CFL reduction: with a device attached, the maximum of
Equation~\ref{eq:cfl} is computed by a Kokkos \texttt{parallel\_reduce}
with a max combiner over the device-resident state --- one formula, two
paths, with the per-cell expression identical to the host one --- so
choosing a time step does not pull the field either. Third, the radiation
prefix product: the scan runs on device views of the beam tables and flux
arrays and never pulls a mirror mid-sweep.

\paragraph{Scope.}
This paper covers single-node parallelism only. MPI domain decomposition is
an orthogonal axis --- halo exchange would wrap the same kernels with no
formula changed --- and is deferred, together with its scaling study, to the
third paper of this series.

\providecommand{\todo}[1]{\textbf{[TODO: #1]}}
\section{Code Validation}

The purpose of this section is contract validation rather than an exhaustive
recreation of every classical astrophysical test.  A run is admitted here when
it establishes one of seven contracts: hyperbolic accuracy, source-term and
conservation behavior, electron and radiation closures, independent multi-fluid
motion, curvilinear geometry, an end-to-end physics chain, or backend
equivalence.  Unless stated otherwise, numerical readings are transcribed
from reports archived with the \texttt{v0.1} evidence snapshot.  The
configuration map records the report or test identifier; reports that state a
binary keep that statement.  We do not promote a measurement from an untracked
or unreproducible run to a new claim.  Table~\ref{tab:tables}
catalogs the corresponding composition/configuration families, and
Table~\ref{tab:validation-config} maps the retained validation records to
their numerical settings and execution backend.

\subsection{Metrics, Units, and Configuration Map}

For the Riemann and Sod reconstruction studies, let $q_i$ denote interior
cell-centred samples; for an extruded two-dimensional problem, the field is
first averaged over the invariant direction.  The archived scaled error is the
arithmetic cell mean
\begin{equation}
L_1(q)=\frac{1}{N}\sum_{i=1}^{N}
  \frac{|q_i-q_{\mathrm{exact},i}|}{S_q},
\label{eq:validation-scaled-l1}
\end{equation}
not a geometry-weighted volume integral.  The report-defined scales are
$S_n=n_{\rm left}$, $S_v=\max(\max_i|v_{{\rm exact},i}|,1)$, and
$S_P=\max_iP_{{\rm exact},i}$; $n_{\rm left}=n_{\rm right}$ for Toro~2.
Extruded problems use the same one-dimensional exact profile after transverse
averaging.  The acoustic convergence study instead reports the unscaled density
mean $L_1(\rho)=N^{-1}\sum_i|\rho_i-\rho_{{\rm exact},i}|$ at the actual final
time.  Its fitted order is the negative slope of a least-squares fit of
$\log e$ versus $\log N$, and a consecutive refinement order is
$\log(e_i/e_{i+1})/\log(N_{i+1}/N_i)$.  Thus ``scaled $L_1$'' labels only the
Riemann/Sod tables; the acoustic table is absolute error in code units.

The dustybox diagnostic is
$L_1=N^{-1}\sum_i|\Delta u_{{\rm num},i}-\Delta u_{{\rm ana},i}|$, with
$L_\infty=\max_i$ of the same cell errors.  Because the box is uniform, every
cell error is identical and $L_1=L_\infty$.  Relative prose comparisons use
the displayed analytic $\Delta u$ at the sampled time.  Conservation drifts
are defined by their archived tests: nucleus drift is
$|N_X(t)-N_X(0)|/N_X(0)$; quasi-neutrality is
$\max|n_e-(n_{\rm HII}+n_{\rm HEII})|/n_e$; and four-fluid momentum drift is
$|p(t)-p(0)|/\max(|p(0)|,10^{-30})$.  In the curvilinear closed-box table,
mass drift is the absolute extrema of the summed mass history, whereas energy
drift is $|E-E_0|/|E_0|$.

For an escape chain,
$\dot M=\sum_s4\pi m_s n_s r^2v_{r,s}$ in CGS units.  Its reported
variability is the population standard deviation of $\dot M$ over cell-centred
samples with $r>4.5R_p$, divided by
$\max(|\langle\dot M\rangle|,10^{-300})$.  Cloud statistics use the archived
cloud-cell mask $\rho_B>1$: mean and maximum are taken over
$|\Delta\mathbf u|=|\mathbf u_A-\mathbf u_B|$ in those cells, and
$v_{B,\max}$ is the maximum Euclidean fluid-B speed in the same mask.  Code
units are identified in Table~\ref{tab:validation-config}; the three escape
chains quote CGS densities, velocities, pressures, masses, radii, and mass
fluxes.  Backend final-time and time-step fingerprints are exact printed
numerical identifiers, not timing or throughput measurements.

\begin{table*}[t]
\centering
\caption{Configuration map of retained validation evidence. Family names refer
to the corresponding report/config directories under \texttt{examples/} and
\texttt{utils/convergence/}.  The evidence column gives the archived report or
test identifier; it does not invent binary provenance that the report does not
state.}
\label{tab:validation-config}
\fontsize{7.4}{8.9}\selectfont
\setlength{\tabcolsep}{3pt}
\begin{tabular}{@{}p{0.125\textwidth}p{0.175\textwidth}p{0.20\textwidth}p{0.235\textwidth}p{0.185\textwidth}@{}}
\hline
Test / family & Purpose & Solver, reconstruction, RK & Geometry, grid, boundaries & Evidence / backend \\
\hline
Hyperbolic accuracy & Riemann exact errors; acoustic order; shock reconstruction & Riemann: HLL/HLLC/TVDLF, PCM--RK1. Acoustic: HLLC with matched PCM/PLM/WENO3--RK. Sod: HLL, three reconstructions, RK1 & Cartesian. Riemann: 1D $N=256$ and extruded $128\times48$, $48\times128$. Acoustic: periodic $N=32$--512. Sod: $N=64$--256 & \texttt{hd\_eval} \S2; \texttt{convergence} \S1--4; \texttt{flux\_eval} \S1--3; Serial \\
Dustybox & Analytic drag relaxation; splitting error & HLLC; PLM--RK2; CFL 0.4 & Cartesian unit interval, periodic; $N=64$, 128; weak/mid/strong $\tau$ & \texttt{dustybox\_1d} \S1--4; Serial \\
Conservation / chemistry & Nuclei, charge, momentum bookkeeping & Nominal HLL--PCM--RK1; stationary uniform state & Cartesian 16-cell box; reflect or copy by contract; 20 steps, $\Delta t\le0.5$\,s & \texttt{test\_a2\_conservation}; v0.1 release suite; Serial \\
Electron / radiation & Closure, partition, antisymmetry, scan limits & Test-specific; tube contracts inherit HLL--PCM--RK1 & Synthetic fields and short tubes defined by tests & \texttt{test\_pe}, \texttt{test\_dpe}, \texttt{test\_shock\_heat}, \texttt{test\_scan}, factory tests; v0.1 suite; Serial where setup is used \\
Multi-fluid functionality & Cloud--shock response; jet and 3D capability & Cloud: HLLC--PLM--RK2. Jet: HLLC--WENO3--RK2. Blast: TVDLF--PCM--RK1 & Cloud: $256\times128$, copy. Jet: $192\times128$, user inlet plus copy. Blast: $64^3$, copy & \texttt{mf\_cloud\_2d}; \texttt{mf\_jet\_2d}; \texttt{mf\_blast\_3d}; Serial \\
Curvilinear geometry & Closed-box conservation; polar defect & HLLC; PCM--RK1; CFL 0.3 & Cylindrical $96\times192$ and spherical $96\times48$, reflect-closed boxes & \texttt{hd\_eval} \S5; Serial \\
End-to-end chains & Tables, scan, chemistry, $P_e$, accounting & HLL; PLM--RK2; CFL 0.4 & Spherical $r\in[R_p,10R_p]$, log grid, $N=128$, 53 beams, user base & \texttt{rad\_hd\_1d}; \texttt{rad\_h\_pe\_1d}; \texttt{rad\_hhe\_pe\_1d}; Serial \\
Backend equivalence & Release suite and backend fingerprints & Contract-specific; probe HLL--PCM--RK1 & Test-specific; probe $512\times256$, reflect/copy, 30 timed steps after warm-up & 394-test report; fingerprint report, \texttt{v0.1}; OpenMP + CUDA builds \\
\hline
\end{tabular}
\end{table*}

\subsection{Hyperbolic Accuracy}

We first test the hyperbolic operator with the source table empty.  Two
one-dimensional Riemann problems are used: Sod's shock tube and Toro problem 2.
Each is also extruded along the second coordinate direction, so that the same
one-dimensional exact solution tests both propagation directions and the
correct treatment of nominally inactive transverse velocities.  The default
cross comparison uses PCM reconstruction and RK1; the convergence study then
matches reconstruction and time integration order.

\begin{table}[t]
\centering
\caption{Representative scaled $L_1$ errors of density for the v0.1
Riemann-solver cross matrix.  The 1D-A row is Sod's problem, 1D-B is Toro
problem 2, and the 2D rows are the corresponding extruded problems.}
\label{tab:validation-riemann}
\begin{tabular}{lcccc}
\hline
Solver & 1D-A & 1D-B & 2D-A & 2D-B \\
\hline
HLL    & $3.515\times10^{-3}$ & $1.237\times10^{-2}$ & $4.962\times10^{-3}$ & $4.628\times10^{-3}$ \\
HLLC   & $3.534\times10^{-3}$ & $1.227\times10^{-2}$ & $5.172\times10^{-3}$ & $4.780\times10^{-3}$ \\
TVDLF  & $3.821\times10^{-3}$ & $1.197\times10^{-2}$ & $4.762\times10^{-3}$ & $4.503\times10^{-3}$ \\
\hline
\end{tabular}
\end{table}

Table~\ref{tab:validation-riemann} shows the expected shock-tube accuracy
level.  HLLC keeps the stationary contact and normal-velocity features sharper
in the Sod family, while TVDLF is more diffusive.  The more difficult Toro
problem has a thin intermediate rarefaction and consequently moves the errors
from the $10^{-3}$ to the $10^{-2}$ range for the first-order baseline.  The
extruded tests also verify that a velocity component which should remain zero
does so and that a genuinely transverse Riemann problem excites the correct
component.

\begin{figure}[t]
\centering
\includegraphics[width=0.98\textwidth]{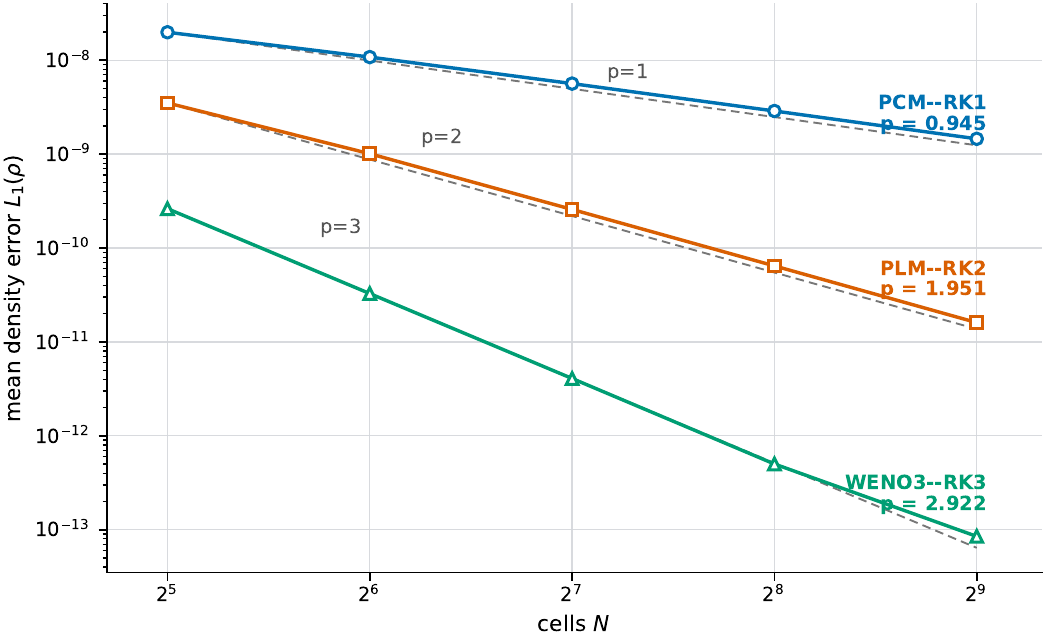}
\caption{Archived acoustic-wave convergence on the periodic unit interval at
one acoustic crossing time.  Each marker is the unscaled mean density error
$L_1(\rho)$ reported at that resolution; dashed grey lines are first-, second-,
and third-order slope guides, and the direct labels give the archived global
least-squares orders.  The runs use HLLC, CFL $0.4$, $N=32$--$512$, the matched
PCM--RK1, PLM--RK2, and WENO3--RK3 combinations, and the Serial backend.  The
archived convergence report is the source; no simulation was rerun for
this figure.}
\label{fig:validation-convergence}
\end{figure}

Reconstruction and time integration are checked with a small-amplitude,
right-propagating acoustic wave run for one acoustic crossing time at
$N=32$, 64, 128, 256, and 512 with CFL number 0.4.
Figure~\ref{fig:validation-convergence} plots every archived density-error
reading and labels the global fitted order for each scheme.  The PCM--RK1
successive refinements rise from 0.88 to 0.98, so the archived report
classifies that baseline as asymptotically first order; PLM--RK2 and
WENO3--RK3 recover their designed smooth-solution orders.

For the Sod problem at $N=256$, PLM reduces the scaled density error from
$7.743\times10^{-3}$ with PCM to $2.035\times10^{-3}$; WENO3 gives
$2.295\times10^{-3}$.  This ordering is expected for a discontinuous solution:
higher-order reconstruction sharpens contacts, but it does not restore a
higher convergence rate at a shock.  RK3 likewise does not by itself improve
this first-order spatial test.  These observations delimit the intended use of
the reconstruction options; they are not a claim that one solver is universally
superior.

\subsection{Source Terms and Conservation Regressions}

\begin{figure}[t]
\centering
\includegraphics[width=0.98\textwidth]{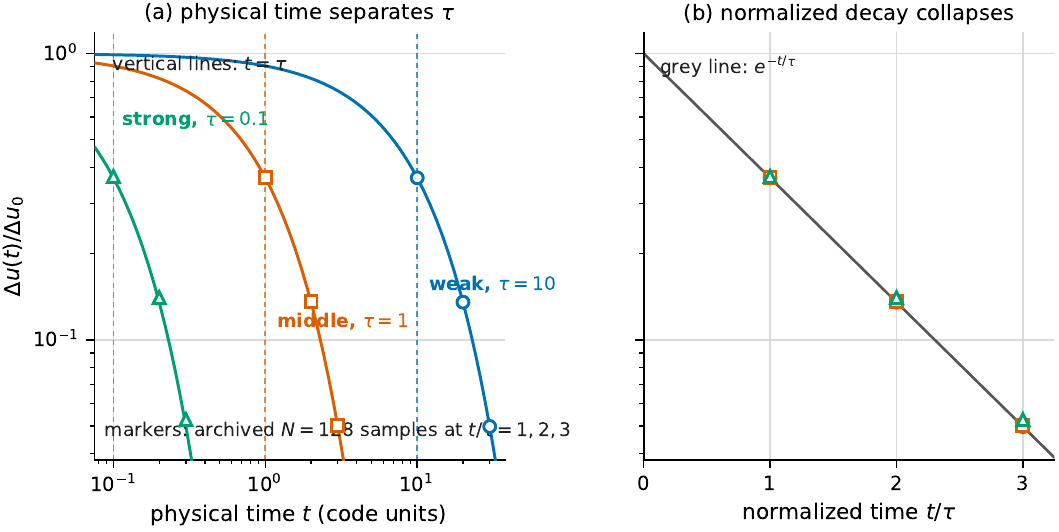}
\caption{Dustybox drag relaxation shown in physical and normalized time.
(a) Physical time separates the weak, middle, and strong drag timescales
$\tau=10$, $1$, and $0.1$; colored lines are the analytic laws and vertical
dashed lines mark $t=\tau$.  (b) The same archived $N=128$ samples are plotted
against $t/\tau$ with the grey analytic exponential, showing their collapse onto
the common decay law.  Markers denote the archived samples at $t/\tau=1,2,3$.
The configuration is a uniform periodic unit box, HLLC--PLM--RK2, CFL $0.4$,
$N=128$, and the Serial backend.  The close agreement exercises the combined
operator-split and implicit-source response; the uniform problem does not
isolate the internal accuracy of the dense LU solve or test spatial
discretization.}

\label{fig:validation-dustybox}
\end{figure}

The first multi-fluid source contract is the analytic ``dustybox'' relaxation
problem.  Two fluids occupy a uniform periodic box with no pressure gradient.
Figure~\ref{fig:validation-dustybox} compares the archived $N=128$
samples with the analytic decay.  Their initial velocity difference decays
analytically as
$\Delta u(t)=\Delta u_0\exp(-t/\tau)$, where $\tau$ is the drag time.  The
table-driven collision band, operator split, and implicit local source solve
are exercised together; the hyperbolic flux is identically zero for this
uniform state.

\begin{table}[t]
\centering
\caption{Dustybox error at $t=3\tau$.  The resolution increase from 64 to 128
halves the error in all three coupling regimes.}
\label{tab:validation-dustybox}
\begin{tabular}{lccc}
\hline
Coupling & $L_1(N=64)$ & $L_1(N=128)$ & Ratio \\
\hline
weak   & $6.488\times10^{-5}$ & $3.244\times10^{-5}$ & 2.00 \\
middle & $6.486\times10^{-4}$ & $3.247\times10^{-4}$ & 2.00 \\
strong & $6.443\times10^{-3}$ & $3.260\times10^{-3}$ & 1.98 \\
\hline
\end{tabular}
\end{table}

The observed first-order behavior is split error: under the CFL condition the
time step scales with cell size, and the Lie-type split is first order.  The
uniform dustybox problem deliberately isolates this combined operator behavior;
it cannot separately resolve the internal accuracy of the dense LU solve.  The
strong-coupling case has $L_1=3.260\times10^{-3}$ at $N=128$, or about 6.6\%
relative error, and should be read as the splitting-dominated upper bound
reported by the archived run.

Conservation and charge contracts are checked independently.  For the hydrogen
and helium network, total H and He nucleus counts have a regression threshold
of $10^{-12}$ relative drift; the archived release run reports bitwise-zero
drift from its equilibrium initialization.  Quasi-neutrality is tested by
comparing the derived electron density with the sum of singly ionized species,
again with a $10^{-12}$ threshold.  With only collision, charge-exchange, and
Coulomb terms enabled, four-fluid total momentum has a $10^{-9}$ threshold and
a measured background drift of $1.5\times10^{-11}$, caused by roundoff in the
velocity-slot update.  We also retain the known saturated pseudo-drift of
order $4\times10^{-5}$ for an initial state away from pressure equilibrium.
It does not grow with step count and is disclosed rather than hidden.

\subsection{Electron Closure and Radiation Contracts}

The electron and radiation tests are contract tests because their purpose is
to establish that the declared options are implemented and reversible, not to
choose a physical truth.  Shock heating is exercised with
distinct ion/electron partition options.  These options are deliberately
reported as model choices; none is treated as the true shock partition.

The quasi-neutral closure regressions check that electrons carry no independent
momentum slot, that the derived electron density and the transported pressure
$P_e$ are readable through the production interface, and that charge weighting
enters the fluid system in the
form specified in Section~3.  The Coulomb, charge-exchange, and electron-energy
channels are checked for exact antisymmetry where antisymmetry is part of the
model contract.

The \texttt{RadiationScan} tests check four analytic limits: a vacuum column
preserves the incident flux; uniform opacity produces exponential attenuation;
an optically thick cell conserves the photon population absorbed from the ray;
and independent rays do not exchange photons.  The end-to-end chains in
Section~4.6 then exercise the same scan with the full 53-ray table rather than
as an isolated kernel.

\begin{figure}[t]
\centering
\includegraphics[width=0.98\textwidth]{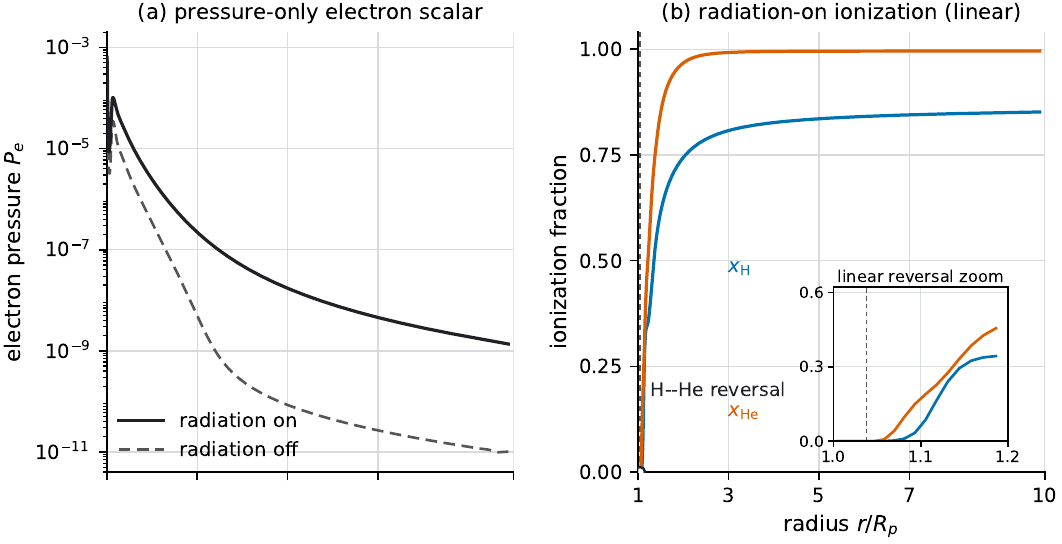}
\caption{Direct pressure-only electron and radiation contract profiles from
the archived \texttt{rad\_hhe\_pe\_1d} final states.  (a) The transported
electron pressure $P_e$ in CGS units with radiation on and off.  (b)
Radiation-on hydrogen and helium ionization fractions, formed only from the
tabulated neutral and ion densities, are shown on a linear axis; the zoom makes
the H--He ionization reversal near the inner boundary visible.  Curves use all
128 archived cell-centred samples without smoothing or resampling.  The runs
use the HLL--PLM--RK2 spherical-chain configuration cataloged in
Table~\ref{tab:validation-config}.  No shock-tier value is derived, and the
plot makes no claim of a physically prescribed electron-heating partition or a
new escape-science result.}

\label{fig:validation-contracts}
\end{figure}

Figure~\ref{fig:validation-contracts} shows the direct final-state profiles
from the archived four-fluid H/He chain with the radiation scan enabled and
disabled.  Panel~(a) is the transported electron scalar $P_e$ in CGS units;
panel~(b) shows the two ionization fractions formed directly from the tabulated
neutral and ion populations on a linear axis.  Their ordering reverses near
$r/R_p\simeq1.04$; the zoom makes that direct contract readout visible.
These are implementation-contract readouts: they
demonstrate that $P_e$ and the declared radiation path are readable end to end,
but they do not select a shock-heating partition or constitute a new
escape-physics prediction.

\subsection{Multi-Fluid Functionality}

The multi-fluid functionality tests emphasize a property that a single-fluid
solver cannot represent: two velocity fields may coexist and then couple
through a declared source band without changing the kernel.  The quantitative
analytic drag reference is the dustybox problem in Section~4.2.  The
two-dimensional cloud--shock interaction supplies the complementary morphological
test.

\begin{figure*}[t]
\centering
\includegraphics[width=0.98\textwidth]{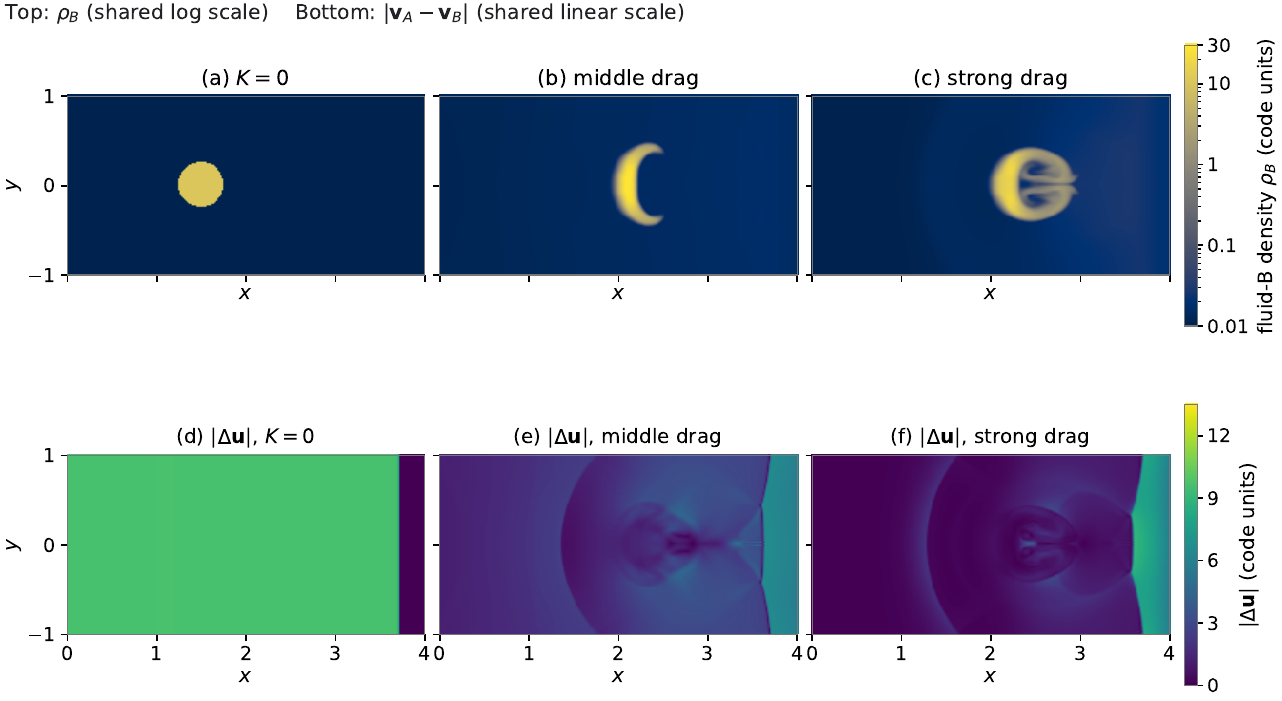}
\caption{Two-fluid cloud--shock morphology at $t=3t_{\rm cc}$.  Top panels show
fluid-B density $\rho_B$ on a shared logarithmic color scale; bottom panels show
$|\Delta\mathbf u|=|\mathbf v_A-\mathbf v_B|$ on a shared linear scale.  Panels
use the archived $256\times128$ cell-centred fields directly, without smoothing
or resampling, for the $K=0$, middle-drag ($K=3.122$), and strong-drag
($K=31.22$) cases.  The numerical settings are cataloged in
Table~\ref{tab:validation-config}.  This is a functional
morphology demonstration, not a convergence or performance claim.}
\label{fig:validation-cloud}
\end{figure*}

The cloud problem uses a Mach 10 shock in fluid A and a dense cloud carried by
fluid B; the cloud has density contrast 10, radius 0.25, and pressure balance
with its own background.  Figure~\ref{fig:validation-cloud} shows the
archived fields, and Table~\ref{tab:validation-cloud} reports
cloud-interior statistics at $3t_{\mathrm{cc}}$, where $t_{\mathrm{cc}}$ is
the cloud-crushing time.  The solver, grid, boundary, and backend settings are
cataloged in Table~\ref{tab:validation-config}.

\begin{table}[t]
\centering
\caption{Two-fluid cloud--shock coupling at $256\times128$ and
$t=3t_{\mathrm{cc}}$.  Cloud cells are defined by $\rho_B>1$; the mean and
maximum are computed over those cells.}
\label{tab:validation-cloud}
\begin{tabular}{lcccc}
Case & $K$ & Mean $|\Delta\mathbf u|$ & Max $|\Delta\mathbf u|$ & $v_{B,\max}$ \\
\hline
off     & 0      & 9.585 & 9.592 & 0.000 \\
middle  & 3.122  & 1.763 & 2.546 & 9.083 \\
strong  & 31.22  & 0.527 & 1.334 & 12.752 \\
\hline
\end{tabular}
\end{table}

With no collision band, the shock and contact pass through fluid A while fluid
B remains exactly stationary in the reported run.  Increasing the drag first
produces a decelerating, stripping cloud and then a strongly locked state with
a cloud-interior velocity difference much smaller than the shock speed.  The
reported maximum fluid-B velocity in the strong case exceeds the incident
post-shock speed; the archived report attributes this to pressure-driven
reacceleration of the non-equilibrium mixture while momentum exchange remains
pairwise antisymmetric.  This is a functional demonstration, not a convergence
study.  WENO3 fails the reported destructive-cloud configuration without a
positivity guard, so PLM is retained and the failure is documented as a recipe
boundary.

The empty-source two-fluid jet and the three-dimensional two-fluid blast are
retained as capability demonstrations: they exercise independent momenta and
dimensionality without assigning them a numerical accuracy claim in this draft.

\subsection{Curvilinear Geometry}

Curvilinear evidence is deliberately separated into three categories.  The
closed-box readings below are conservation bookkeeping for geometry-weighted
sums.  The G-CYL point-explosion run is retained only as a
cylindrical-coordinate capability demonstration; it is not evidence of perfect
circular symmetry.  The spherical-polar G-SPH point-explosion panels are
excluded from spherical-symmetry validation because the polar geometric source
defect remains unresolved in v0.1.

\begin{table}[t]
\centering
\caption{Closed-box conservation extrema in the archived curvilinear geometry
runs.  Mass drift is absolute; energy drift is relative to $E_0$.}
\label{tab:validation-geometry-conservation}
\begin{tabular}{lcc}
\hline
Geometry & Maximum mass drift & Maximum energy drift \\
\hline
Cylindrical $(R,z)$ & $5.551\times10^{-17}$ & $2.767\times10^{-15}$ \\
Spherical $(r,\theta)$ & $5.551\times10^{-17}$ & $4.309\times10^{-16}$ \\
\hline
\end{tabular}
\end{table}

The mass variation is about four ULP and is caused by summation-order changes
in the unpack--reshape--sum path, while the energy drift remains below
$3\times10^{-15}$ in both geometries.  Reflecting walls impart momentum, so
total momentum is intentionally not asserted in these closed boxes.  These
closed-box readings are conservation bookkeeping only; they do not turn either
point-explosion run into a symmetry-convergence test.

In contrast, the spherical-polar diagnostic reveals a real defect near the
pole: the $\theta$-momentum update uses an area-weighted finite-volume
divergence without the compensating pressure geometric source.  The defect is
unresolved in v0.1.  The archived probe reports maximum $|v_\theta|=0.50$,
0.85, 1.25, and 1.48 for $n_\theta=8$, 16, 32, and 48; the error grows rather
than converges.  The archived spherical point-explosion panels are therefore
excluded as Paper~1 symmetry evidence.  A one-dimensional radial formulation
is a separate radial-geometry check and does not exercise the polar update;
it does not repair the polar defect.

Paper~1 therefore includes no curvilinear point-explosion figure.  The
closed-box table remains conservation bookkeeping, and the geometry correction
belongs to a later release rather than to a reinterpretation of the archived
panels.

\subsection{End-to-End Application Chains}

Three planetary-escape chains connect the Python declaration, table assembly,
hyperbolic transport, local implicit chemistry and collision sources, electron
pressure, and ray scan.  They are used as accounting checks of the complete
chain.  They do not rerun or revise the scientific conclusions of any previous
escape simulation.

\begin{table*}[t]
\centering
\caption{End-to-end chain readings from the archived reports.  Variability is
the population standard deviation of $\dot M$ over cell-centred samples with
$r>4.5R_p$, divided by the magnitude of their mean.}
\label{tab:validation-chain}
\begin{tabular}{p{0.16\textwidth}p{0.24\textwidth}p{0.32\textwidth}p{0.10\textwidth}}
\hline
Chain & Physics retained & Representative reading & Variability \\
\hline
\texttt{rad\_hd\_1d} & HI/HII, 53 rays & $x_{\mathrm{ion}}=0.973$ at outer boundary; $\dot M=9.35\times10^9$\,g\,s$^{-1}$ & $8.0\times10^{-4}$ \\
\texttt{rad\_h\_pe\_1d} & H ions, electron pressure, Coulomb & $\dot M=2.38\times10^{10}$\,g\,s$^{-1}$ & 0.017 \\
\texttt{rad\_hhe\_pe\_1d} & four species, 53 rays, 26 bands, $N=21$ & $\dot M=3.499\times10^9$\,g\,s$^{-1}$ & 0.071 \\
\hline
\end{tabular}
\end{table*}

The three cases show that changing the component and source tables changes the
resolved chain while the kernel remains fixed.  The full H/He case also
demonstrates that the declared factory produces the intended four-fluid state,
dual absorbers, and dense $21\times21$ source system.  Its time step is capped
at 10\,s for source-step stability; any discussion of physical elapsed time
must state that cap.  These are reproducibility and accounting readings, not
new scientific escape-rate predictions.

\subsection{Backend Equivalence}

The release report records 394 passing tests for the complete suite used at
v0.1.  Backend validation has two layers.  First, Serial, OpenMP, and CUDA
regression checks compare pinned quantities, but their acceptance criteria are
test- and quantity-specific: empty-source Sod scalar identifiers are exact
pins, selected field summaries use prescribed ULP-scale tolerances for
platform mathematical-library roundoff, and production-like fingerprints are
archived probes rather than a blanket bitwise guarantee for every field.
Second, physics invariants are checked on every backend so that a scheduling
change cannot silently alter source antisymmetry or radiation bookkeeping.

The end-to-end fingerprint quoted here is the archived probe in which the
OpenMP and CUDA executables report the same final time,
$t=0.032679101492116794$, and the same final time step,
$\Delta t=0.0010109037898408007$.  This is an exact printed-identifier match
for that probe.  It is not a claim that every field is bitwise identical
across the backends, and it is not a timing measurement.

The v0.1 handoff does not freeze a complete hardware, compiler, Kokkos, and
thread-affinity manifest for these checks.  The OpenMP regression recorded in
that handoff also remains unresolved.  Paper~1 therefore makes no speedup,
parallel-efficiency, or throughput claim, interprets no isolated CUDA
wall-clock observation, and excludes MPI scaling.  A normalized single-node
benchmark is deferred until the benchmark environment, workload definition,
and backend status are frozen.

\section{Summary and Outlook}

Xenia is a table-driven multi-fluid hydrodynamics framework: fluid species,
collision bands, chemical and thermal source terms, and radiation channels are
declared in Python, while a versioned ABI serves the same C/Kokkos kernel
implementation in separate Serial, OpenMP, and CUDA backend builds.
Its initial application
surface is planetary atmospheric escape, but HD~209458b is only the initial
component table rather than a specialization compiled into the solver.
The practical value is a lower modeling barrier: changing species
definitions, collision bands, chemistry, heating and cooling, or radiation
channels changes versioned Python tables rather than kernel code, and the
same kernel evaluates the resulting tapes.

This paper contributes three methods.  The table-driven multi-fluid framework
assembles pairwise species coupling from channel tables into $\mathcal L_c$:
changing species definitions, collision bands, chemistry, heating and cooling,
or radiation channels changes versioned data crossing the ABI, not the generic
kernel implementation.  The pressure-only quasi-neutral electron closure keeps
neutral and ion species as independent dynamical constituents while fixing the
electron density by quasi-neutrality, advancing no electron mass or momentum,
transporting only $P_e$, coupling ions through charge-weighted
electron-pressure work, and exposing shock-energy partition as explicit model
options.  The operator-split stiff-source treatment sequences nonlocal
\texttt{RadiationScan}, the hyperbolic operator $L_f$, and a cell-local
implicit source solve; $F_b$ is frozen after the scan, the same symbolic tapes
provide an analytic finite-difference-free Jacobian, and each cell uses a
direct dense LU solve.  The Python \texttt{setup}, \texttt{step},
\texttt{fields}, and \texttt{close} workflow, schema versioning, Kokkos
deployment, and release infrastructure are supporting implementation choices
rather than a fourth physics or methods contribution.  Paper~1 reports
functional backend deployment and
backend-equivalence fingerprints, not backend speed; normalized performance
and distributed scaling remain deferred.

Several boundaries are equally part of the contribution.  Conservation and
positivity are not claimed to match the mature implicit-drag and
space-plasma machinery of older production codes; they are controlled by
regressions and disclosed failure boundaries.  Shock partition between
electron and ion heating is not prescribed by the underlying physics,
so Xenia offers explicit options rather than declaring one
partition the truth.  The planetary-escape chains
are reproducibility and accounting tests and do not redraw the scientific
escape conclusions of earlier simulations.  Finally, the radiation operator is
a radial Beer--Lambert prefix product, not a general radiation solver.

The source tree has already absorbed an ideal-MHD superset.  With magnetic
fields disabled, the implementation returns to the HD path under a locked
zero-field regression; magnetic physics and its dedicated validation are
reserved for Paper~2.  This separation lets the present paper establish the
HD, table, closure, and splitting contracts without borrowing capabilities
that belong to the next paper.

Paper~2 will describe ideal MHD with constrained transport, traced radiation,
and the additional dissipation channels enabled by that formulation.  Paper~3
will address distributed-domain decomposition, MPI scaling, and adaptive mesh
refinement.  Together, the three papers separate the portable multi-fluid HD
foundation, its magnetic extension, and its scalable domain decomposition
into independently testable layers.

The public Xenia code release is available at
\url{https://github.com/YFG-tec/xenia}, tagged
\texttt{v0.1.1-paper.1}.  Xenia is distributed under GPLv2; Kokkos remains a
separate Apache-2.0 submodule and is not subsumed by that grant.  The numerical evidence uses the archived v0.1 evidence snapshot; the public tag
is a cleaned release assembled for this code paper.

\section*{Acknowledgments}
After graduating from Yunnan Astronomical Observatory, the author joined Beijing ENN Fusion Energy Science and Technology Co., Ltd. (formerly the ENN Fusion Energy Research Institute), where work on fusion as a clean and sustainable source of energy opened a broader scientific and technical perspective. The author thanks the members of the Intelligent Capability Group for their generous support during this transition, and is especially grateful to Dr.~Tianyuan Liu for his academic guidance and his thoughtful care in both work and daily life. Finally, the author wishes ENN every success in becoming a pioneer in the commercialization of hydrogen--boron fusion energy.

\appendix
\section{Reproducibility and User Path}
\label{app:reproducibility}

This appendix gives the minimal user path, the supported build and test
sequence, and the provenance boundary for the numerical evidence and figures.
It is intended to let a reader distinguish what is regenerated from archived
release evidence from what still requires human-supplied public provenance.

\subsection{Minimal User Lifecycle}
\label{app:user-lifecycle}

The Python front end exposes a compact four-call lifecycle:
\begin{verbatim}
import xmf

run = xmf.setup(xmf.shock_tube())
while run.t < run.tstop:
    run.step()
hi = run.fields["HI"]
print(run.t, hi.n.shape, float(hi.n.max()))
run.close()
\end{verbatim}
Here \texttt{setup()} packs the Python problem declaration into the versioned
ABI, \texttt{step()} advances the operator sequence, \texttt{fields} reads the
interior state, and \texttt{close()} releases the kernel resources.
The built-in \texttt{shock\_tube()} configuration is only a minimal one-fluid
demonstration.  Realistic multi-fluid problems replace its Python dictionary
with the appropriate fluid, source, chemistry, radiation, grid, and numerical
settings; those declarations are compiled to tapes rather than used to select a
species-specific kernel.

\subsection{Build, Synchronization, and Tests}
\label{app:build-test}

The package requires Python~$\geq 3.10$ and NumPy at runtime; pytest and
matplotlib are developer extras.  The orchestration and mesh layers are C99,
while the Kokkos sweep and source layers are C++17.  scikit-build-core supplies
the CMake/Ninja build path, and Kokkos~4.7 is vendored with the source tree.
Serial, OpenMP, and CUDA are separate compile-time backend builds: one compiled
library exposes one backend, and there is no runtime backend switching.

A platform-neutral source checkout is built and checked with
\begin{verbatim}
python -m venv .venv
<activate the selected interpreter>
python -m pip install --upgrade pip
python -m pip install .[dev]
python scripts/sync_xmf.py
python -m pytest tests -q
\end{verbatim}
The interpreter may be activated or invoked by its explicit path.  After source
edits, \texttt{scripts/sync\_xmf.py} is the supported path for rebuilding and
installing the kernel into the live \texttt{xmf} import tree.  Running tests
against an unsynchronized tree can test a stale library rather than the newly
built binary.

The exact hardware, compiler, Kokkos, and backend environment used for a
released run belongs to its version/environment manifest.  That manifest has not
been frozen for public release here and is therefore not reconstructed in this
paper.

\subsection{Release, Data, and Figure Provenance}
\label{app:provenance}

The numerical evidence in this paper was produced with the archived v0.1
evidence snapshot.  Validation quotes numbers from the archived reports and
tests identified by name in the configuration map; this paper does not trace
those records through internal development commits.

The figure-generation scripts are part of the paper source.  The validation
scripts read archived reports and committed field outputs directly; the concept
diagrams are rendered by their own deterministic scripts.  Regenerating these
figures does not rerun the simulations.  No new timing benchmark or
performance dataset is claimed.  The known G-SPH defect panels are not used
as validation evidence.

\subsection{Software Availability Boundary}
\label{app:software-availability}

The public source release is
\url{https://github.com/YFG-tec/xenia}, tagged
\texttt{v0.1.1-paper.1}.  It contains the Python semantic layer, the
C/Kokkos kernel sources, tests, examples, public documentation, build files,
and the Kokkos submodule pointer.  Xenia is released under GPLv2; Kokkos is
licensed separately under Apache 2.0 with the LLVM exception.

The public tag is a cleaned release assembled for this code paper, whereas the
numerical readings use the archived v0.1 evidence snapshot.  The release
therefore provides the implementation and example surface, while the paper
identifies numerical evidence by archived report and test names.  The arXiv
preprint identifier will link back to this GitHub release when the preprint is
posted.

\end{document}